\documentclass[preprint,preprintnumbers,superscriptaddress,amsmath]{revtex4}
\usepackage{url}
\usepackage{epsfig}
\usepackage{graphicx}% Include figure files
\usepackage{epstopdf}
\usepackage[psamsfonts]{amssymb}
\usepackage{amsmath}
\usepackage{indentfirst}

\begin{document}

\title{Nonlinear theory and tests of earthquake recurrence times}

\author{D. Sornette}
\affiliation{Department of Management, Technology and Economics,
ETH Zurich, Kreuzplatz 5, CH-8032 Zurich, Switzerland}
\affiliation{Institute of Geophysics and Planetary Physics
and Department of Earth and Space Sciences,
University of California, Los Angeles, CA 90095}
\email{dsornette@ethz.ch}

\author{S. Utkin}
\affiliation{Mathematical Department,
Nizhny Novgorod State University, Gagarin prosp. 23,
Nizhny Novgorod, 603950, Russia}
\email{sergei_utkin@mail.ru}

\author{A. Saichev}
\affiliation{Department of Management, Technology and Economics,
ETH Zurich, Kreuzplatz 5, CH-8032 Zurich, Switzerland}
\affiliation{Mathematical Department,
Nizhny Novgorod State University, Gagarin prosp. 23,
Nizhny Novgorod, 603950, Russia}
\email{saichev@hotmail.com}

\begin{abstract}
We develop an efficient numerical scheme to solve accurately the
set of nonlinear integral equations derived previously in (Saichev and Sornette, 2007), 
which describes the distribution of
inter-event times in the framework of a general model of earthquake clustering
with long memory. Detailed comparisons between the linear and nonlinear versions of the theory
and direct synthetic catalogs show that  the nonlinear theory provides an excellent fit to the synthetic catalogs, while there are significant biases resulting from the use of the linear approximation. We then address the suggestions proposed by some authors
to use the empirical distribution of inter-event times to obtain a better determination
of the so-called clustering parameter. Our theory and tests against synthetic and
empirical catalogs find a rather dramatic lack of power for the distribution of inter-event times to
distinguish between quite different sets of parameters, casting doubt on the
usefulness of this statistics for the specific purpose of identifying the clustering parameter.
\end{abstract}

\date{\today}

\maketitle
%\fontsize{14pt}{24pt}\selectfont

\section{Introduction}

Most complex systems of interest in the natural and social sciences
exhibit intermittent bursts of activity interspersed within long times of
reduced activity. A simple metric to characterize this property consists
in the distribution of recurrence (also called ``waiting'' or
``inter-event'') times between (suitably defined) events.
Recently, the literature has undergone itself a burst of publication activity on
this topic, motivated by the idea that distributions of recurrence times
may be one of the most important complexity measures for both random
fields and nonlinear dynamical systems \cite{Gao}. The applications
include recurrence time and anomalous transport \cite{Zaslavsky91},
waiting times between earthquakes \cite{Corral03,Corral2004a,Livina,SaichevSor06,SaichevSor07}
and rock fractures \cite{Davisenetal07},
time intervals between consecutive e-mails \cite{Barabasi_Nature05} and between
web browsing, library visits and stock trading \cite{Vasquez_et_al_06}.

Much of the recent interest of the statistical physics community focused on applying
scaling techniques, which are common tools in the study of critical phenomena, to
the statistics of inter-earthquake recurrence times or waiting times
\cite{Baketal02,Corral03,Corral2004a,Corral2004b,Corral2005a,Corral2005b,Corral2006,Corral_Christensen06,DavidsenGoltz04,Livina06}. Many of the claims made in these recent articles on recurrence statistics have either been challenged, refuted or explained by previously known facts about earthquake
statistics \cite{Lindmanetal05,Lindmanetal06,Molchan05,Hainzl2006,SaichevSor06,SaichevSor07}.
In particular, two of us \cite{SaichevSor06,SaichevSor07} have developed a
general theory of the statistics of inter-event times in the framework of the
general class of self-excited Hawkes conditional Poisson processes
\cite{Hawkes71a,Hawkes71b,HawkesOakes74} adapted to modeling
seismicity. The corresponding model is known as the epidemic-type aftershock sequence
(ETAS) model, in which any earthquake may trigger other earthquakes, which in turn
may trigger more, and so on. Introduced in slightly different forms by Kagan and
Knopoff \cite{KK81} and Ogata \cite{Ogata88}, the model describes statistically the spatio-temporal
clustering of seismicity. Using three well-known statistical laws of statistical
seismicity (the Gutenberg-Richter, the Omori law and the productivity law),
the empirical observations on the distribution of earthquake recurrence times
can be explained within this model without invoking additional mechanisms other than
the well-known fact that earthquakes can trigger other earthquakes  \cite{SaichevSor06,SaichevSor07}.

A recent development is the proposition that inter-event time distributions may
provide a new and more reliable way to measure of the so-called background earthquake activity
\cite{Hainzl2006,Hardebeck07}. This question arises as follows: if earthquakes trigger other
earthquakes, how much of the observed seismicity is due to past seismicity
(endogenous origin) and how much is resulting from an ``external'' driving source
(exogenous origin) often referred to as ``background'' seismicity thought to reflect
the driving tectonic forces at large scales. This question obviously generalizes
to any system in which future events may be in part triggered by past events,
such as in commercial sales \cite{Sornetteetal04} and web browsing activity \cite{Vasquez_et_al_06,CraneSornette07}. Within the ETAS framework,
the fraction of events in a given catalog which have been triggered by previous
events can be shown \cite{Helmsor03} to be nothing but the so-called branching ratio $n$,
defined mathematically as the average number of first-generation events triggered
by a given preceding event \cite{Helmsor02}. Reciprocally, the fraction of
background events is equal to $1-n$ (note that these models assume that
the triggering branching-like processes are sub-critical: $n<1$).
The degree to which the parameter $n$ can be retrieved from the distribution
of inter-event times relies on departure from universality pointed out
by Hainzl et al.\cite{Hainzl2006} and two of us  \cite{SaichevSor06,SaichevSor07}.
In this respect, the ETAS model provides an excellent training ground.
Using synthetic catalogs generated with the ETAS model, Hainzl et al.\cite{Hainzl2006}
found that the estimation of $n$ using the  distribution
of inter-event times  is better than from the application
of a standard declustering procedure \cite{Reasenberg85}.

More progress can be achieved by a better understanding of the
sensitivity of the distribution of inter-event times to the
branching ratio $n$. In principle, the theoretical framework based
on the technique of probability generating functions developed in
Ref. \cite{SaichevSor06,SaichevSor07} provides an ideal approach to
this problem. However, this previous effort was limited on two
accounts. First, while Saichev and Sornette derived the full exact
nonlinear integral equations of the problem, they ended solving
their linearized versions in order to derive the distribution of
inter-event times. The present paper keeps the full nonlinear
integral equations and shows that the linear simplification leads to
systematic biases in the estimations of the key parameters of the
ETAS model, and in particular of the branching ratio $n$ which has
been the focus of recent interest in the seismological community
\cite{Hainzl2006,Hardebeck07}. Secondly, only preliminary sensitivity analysis was
performed with respect to $n$. The present paper presents a detailed
treatment of the full exact nonlinear equations providing the
distribution of inter-event times for the ETAS model and discusses
how well $n$ can be constrained.

The organization of the presentation is as follows. Section 2 describes
the theoretical framework developed by Saichev and Sornette \cite{SaichevSor06,SaichevSor07}
and summarizes their main results, essentially based on a linear
approximation to the full nonlinear equations that they derived. Section 3 focuses on these nonlinear equations and presents the numerical scheme that has been used to solve them.
Detailed comparisons between the linear and nonlinear versions of the theory
and direct ETAS simulated catalogs are presented. With improved adaptive mesh grids,
it is shown that the nonlinear theory provides an excellent fit to the synthetic ETAS
catalogs, while there are significant biases resulting from the use of the linear approximation.
Section 3 concludes by a synthetic test demonstrating the possibility to use the nonlinear
theory to invert for two of the unknown parameters, if constraints exist on the other three
parameters of the model.
Section 4 applies these results to the empirical data set treated by Corral \cite{Corral03}.
We find a rather dramatic lack of power for the distribution of inter-event times to
distinguish between quite different sets of parameters, casting doubt on the
usefulness of this statistics for the specific purpose of identifying the clustering parameter $n$.

\section{Summary of results obtained by Saichev and Sornette \cite{SaichevSor06,SaichevSor07}}

\subsection{The ETAS model}

The ETAS model views the flow of future seismicity as being triggered by past
seismicity and by a few background events. Each earthquake is assumed to have
the potential to trigger future earthquakes according to three laws capturing the nature of seismicity
viewed as a marked point-process. We restrict this study to the temporal domain only,
summing over the whole spatial domain of interest. First, the magnitude of any earthquake,
regardless of time, space or magnitude of the mother shock, is drawn randomly
from the exponential Gutenberg-Richter (GR) law. Its complementary
cumulative probability distribution is expressed as
\begin{equation}
Q(m)=10^{-b(m-m_0)}.\label{eq_g}
\end{equation}
where the constant exponent $b$ is typically close to one, and the cut-off $m_0$
serves to normalize the pdf. We do not consider the influence of an upper cut-off
$m_{\rm max}$, usually estimated in the range $8-9.5$
\cite{Kagan99,Pisaetal07}, because its impact is quite weak in the calculations.

Second, the model assumes that direct aftershocks are distributed in time according
to the modified ``direct'' Omori law (see Ref.\cite{Utsuetal95} and references therein).
Denoting the usual Omori law exponent by $p=1+\theta$ and assuming
$\theta >0$, the
normalized pdf of the Omori law can be written as
\begin{equation}
\Phi(t)=\frac{\theta c^\theta}{(t+c)^{1+\theta}}~,
\label{eq_o}
\end{equation}
where $t$ is the time since the earthquake and $c$ is a regularizing constant preventing
the divergence of the rate at small times.

Third, the number of direct aftershocks of an event of magnitude $m$ is assumed to
follow the productivity law:
\begin{equation}
\rho(m) = \kappa \cdot 10^{\alpha (m-m_0)}~,  ~~~~m_0 \leq m~,
\label{producad}
\end{equation}
where $\kappa$ and $\alpha$ are constants. Note that the productivity law (\ref{producad})
is zero below the cut-off $m_0$, i.e., earthquakes smaller than $m_0$ do not trigger
other earthquakes. The existence of the small-magnitude
cut-off $m_0$ acts as a ``ultra-violet'' cut-off which is necessary to ensure the convergence of the models of triggered seismicity for $\alpha \leq b$.

These laws are combined with the fundamental defining ETAS equation
\begin{equation}
\Lambda(t) = \omega + \sum_{i | t_i<t}  \rho(m_i) \Phi(t-t_i)~,
\label{gnontq}
\end{equation}
giving the conditional Poisson intensity $\Lambda(t)$ for the occurrence of the next
event, conditioned on the history of past events ${\cal H}(t)=\{...(t_i, m_i), ...., (t_1, m_1)\}$.
Here, $t_i <t$ (respectively $m_i$) is the time of occurrence (respectively magnitude)
of the $i$-th earthquake counted from the present time $t$. The term $\omega$
is the background contribution assumed to embody the effect of the large scale
tectonic driving. Taking the expectation of (\ref{gnontq}) yields the average seismic rate
\begin{equation}
{\rm E}[\Lambda(t)] = {\omega \over 1-n}~,
\end{equation}
where $n$ is the key parameter of the ETAS model defined as the number of direct aftershocks
per earthquake, averaged over all magnitudes:
\begin{equation}
n \equiv \int_{m_0}^{+\infty} |dQ(m)/dm| \rho(m) dm = {\kappa b \over (b- \alpha)} ~.
\end{equation}
As recalled in the introduction,
the fraction of events in a given catalog which have been triggered by previous
events can be shown \cite{Helmsor03} to be exactly given by this
``branching ratio'' $n$.

\subsection{Mathematical formulation for the determination of the distribution
of inter-event times}

Saichev and Sornette \cite{SaichevSor06,SaichevSor07} used the formalism
of probability generating functions to calculate from first principles for the ETAS
model the distribution $H(\tau,m)$ of waiting times between events
of magnitudes larger than or equal to $m$ in a region
of seismicity rate $\lambda(m)$. In agreement with previous works
\cite{Corral03,Corral2004a,Livina,SaichevSor06,SaichevSor07}, we express
$H(\tau,m)$ as
\begin{equation}
H(\tau,m) \simeq \lambda(m) f(\lambda(m) \tau)~,
\label{gntpgtpqw}
\end{equation}
so that the dependence on the local seismicity rate is absorbed in the
variable $\lambda(m)$ while the more general functional form is captured
by the function $f(x)$. Saichev and Sornette first used the
general relation
\begin{equation}
H(\tau,m)= {1 \over \lambda(m)} {d^2 P(\tau,m) \over d\tau^2}~,
\label{d2pdf}
\end{equation}
where $P(\tau,m)$ is the probability of absence of events of
magnitude larger than or equal to $m$ within the interval $[t,t+\tau]$.
The following expression was obtained
\begin{equation}
P(\tau,m)=\exp\left( -\int\limits_0^\infty N(t,\tau,m)dt
-\int\limits_0^\tau N_-(\tau',m)d\tau' \right)~,
\label{eq_4}
\end{equation}
with the auxiliary functions $N_-(\tau,m)$ and $N(t,\tau,m)$ given by
the following nonlinear integral implicit equations
\begin{equation}
N_-(\tau,m)=1-\Psi[\Phi(\tau)\otimes N_-(\tau,m)]
+Q(m)\Psi[Q^{-1/\gamma}(m)\Phi(\tau)\otimes N_-(\tau,m)]~ ,\label{eq_1}
\end{equation}
\begin{equation}
N(t,\tau,m)=1-\Psi[\Phi(t)\otimes N(t,\tau,m) +\Phi(t+\tau)\otimes
N_-(\tau,m)]~,
\label{eq_2}
\end{equation}
where
\begin{equation}
\gamma \equiv {b \over \alpha}
\end{equation}
is assumed larger than $1$ (but probably close to $1$).
$N_-(\tau,m)$ and $N(t,\tau,m)$ have the following probabilistic interpretation:
\begin{itemize}
\item $N_-(\tau,m)$ is the probability that either some background earthquake occurs
in the time window  $[t,t+\tau]$ which has a magnitude larger than $m$, or, if its magnitude is smaller than $m$, given that it occurred at time $t$, that it will generate at least one aftershock (or their subsequent daughters) of magnitude larger than $m$ within the interval $[t,t+\tau]$. 
\item Analogously, $N(t,\tau,m)$ is the probability that some background earthquake of
magnitude larger than $m$, occurring at instant $t=0$, will generate at least one
aftershock of magnitude larger than $m$ within the time interval $[t,t+\tau]$.
\end{itemize}
The symbol
$\otimes$ stands for the convolution operation over the variable
$\tau$ in (\ref{eq_1}) and in the second part of the argument of
the function $\Psi$ in (\ref{eq_2}), and over the variable $t$ in
the first part of the argument of the function $\Psi$ in (\ref{eq_2}).
The function $\Psi(z)$ is expressed through the incomplete
Gamma-function:
\begin{equation}
\Psi(z)=\gamma(\kappa z)^\gamma \Gamma(-\gamma,\kappa
z).\label{eq_psi}~.
\end{equation}
For convenience, we use its expansion in powers of $z$:
\begin{equation}
\Psi(z)=1-nz+\beta z^\gamma-\eta z^2+\ldots
\label{eq_3}
\end{equation}
where $\beta$ and $\eta$ are two numerical constants which can be expressed
in terms of $\kappa$ and $\gamma$.

It is clear that the distribution $H(\tau,m)$ of inter-event times
$\tau$ depends on the magnitude cut-off $m$ of events used to
construct this distribution. A natural value for this cut-off is the
magnitude $m_d$ of so-called completeness of the considered catalog,
above which all earthquakes are thought to be recorded by the
existing seismic network. This detection threshold $m_d$ has evolved
over time together with the technology and density of the seismic
networks. In our comparison with Corral's analysis presented below,
we use the values $m_d$ reported by him for each corresponding
catalog.

The goal of this paper sequel is to calculate the full solution of (\ref{eq_1},\ref{eq_2})
leading to the expression of $H(\tau,m)$ given by (\ref{d2pdf}) and to compare
this prediction to the data analysis performed by Corral \cite{Corral03}
in order to bracket the three key parameters of the ETAS model, $\theta, \gamma$ and $n$.
The first one describes the direct Omori law.
The second one, given the well-known $b$-value, provides a new estimate for
the productivity exponent $\alpha$. The third one $n$ is directly associated
with the fundamental question in seismicity of how much clustering occurs
in recorded catalogs, as discussed in the introduction.

\subsection{Analytical solution using the linear approximation}

The determination of the form of $f(x)$ defined in (\ref{gntpgtpqw}) can be
analytically resolved only by reducing equations (\ref{eq_1}) and
(\ref{eq_2}) to their linear approximations, i.e., when only the two first summands of the
expansion~(\ref{eq_3}) are considered:
\begin{equation}
\Psi(z)\cong 1-nz.
\label{gjnbtpqa}
\end{equation}
Using this approximation, Saichev and Sornette
\cite{SaichevSor06,SaichevSor07} introduces for convenience the
auxiliary function $g(\tau,m_d)$ defined by
\begin{equation}
g(\tau,m_d)=1-\frac{N_-(\tau,m_d)}{N_-(m_d)},\quad
N_-(m_d)=\lim_{\tau\rightarrow\infty}(N_-(\tau,m_d)).\label{eq_5}
\end{equation}
With the linear approximation (\ref{gjnbtpqa}), we have
\begin{equation}
N_-(m_d)=\frac{Q(m_d)}{1-\delta},\quad
\delta=n[1-Q^{1-1/\gamma}(m_d)]
\label{eq_6}
\end{equation}
With (\ref{gjnbtpqa}), the main remaining problem of solving equation~(\ref{eq_2})
can be done by representing the first integral in~(\ref{eq_4})  via the second one,
so that one just needs to determine the function $g(\tau,m_d)$, from which
one obtains
\begin{equation}
P(\tau)=\exp\left(-\frac{1-n}{1-\delta}\tau-
\frac{1-n}{\delta}\Delta\int\limits_0^\tau g(\tau',m_d)d\tau'
\right),\label{eq_9}
\end{equation}
where $\Delta=\frac{n}{1-n}-\frac{\delta}{1-\delta}$. Here, we have dropped
the explicit dependence on the magnitude, except in the function $g$ which is written
as dependent on the threshold magnitude $m_d$.
Then, using a quasi-static approximation for $g(\tau,m_d)$, Saichev and Sornette obtained
\begin{equation}
g(\tau,m_d)\cong\frac{\delta a(\tau)}{1-\delta+\delta a(\tau)}~,
\label{eq_q}
\end{equation}
where
\begin{equation}
a(\tau)= \int_\tau^\infty \Phi(t') dt' = {c^\theta \over (\tau + c)^\theta}~ .
\label{at}
\end{equation}
Using the dimensionless variable
$x=\lambda\tau$, where $\lambda=\omega Q(m_d)/(1-\delta)$
and $\omega$ is the seismic rate of spontaneous seismic sources,
expression (\ref{d2pdf}) with (\ref{gntpgtpqw}) yields
\begin{equation}
f(x)=\frac{d^2 \varphi(x,m_d)}{dx^2}~,\quad
\varphi(x,m_d)=P(\tau)=P\left(\frac x\lambda \right) ,
\label{eq_deff}
\end{equation}
leading finally to the dimensionless distribution of inter-event times:\begin{equation}
f(x)=\left(\theta\nu(1-\delta)\epsilon^{-\theta}
(x+\epsilon)^{\theta-1}g^2(x,m_d,\theta)+ [\eta+\nu
g(x,m_d,\theta)]^2 \right) \varphi(x,m_d),\label{eq_10}
\end{equation}
Here,  $\epsilon=\lambda c$, $\nu=(1-n)\Delta$ and
$g(x,m_d,\theta)=g(\tau,m_d)=g(x/\lambda,m_d)$.

\begin{figure}[t]
\includegraphics[width=12cm]{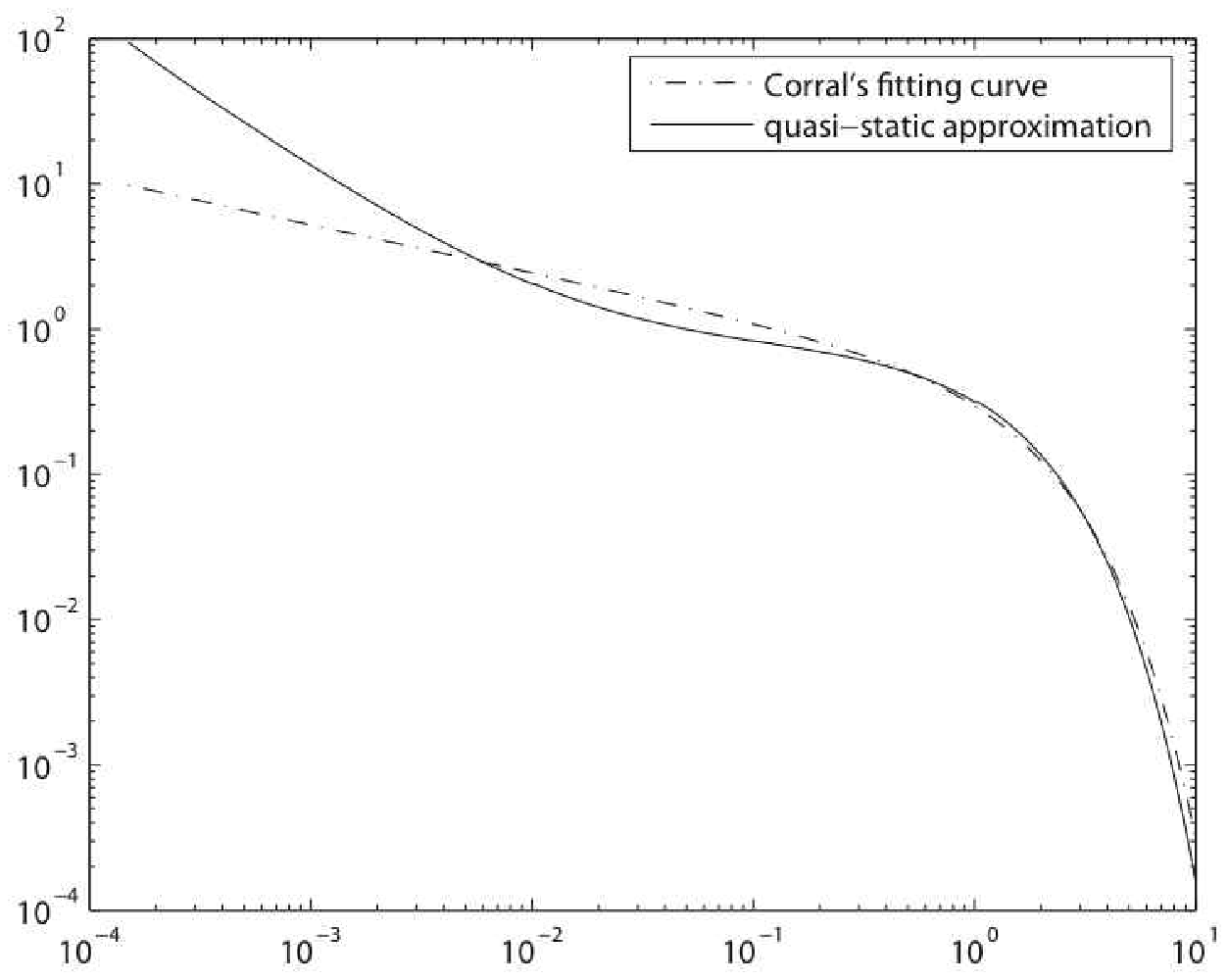}
\caption{Non-dimensional probability density function $f(x)$ of
inter-events times for $m_d-m_0=2, \gamma=1.2, n=0.9, \theta=0.03$
(continuous line) compared with Corral's phenomenological expression
\cite{Corral03} (dashed-dotted line).} \label{fig:quazi}
\end{figure}

Fig.~\ref{fig:quazi} reproduces the comparison obtained
in Ref.\cite{SaichevSor06,SaichevSor07}  between
the function $f(x)$ given by (\ref{eq_10}) and
Corral's phenomenological functional fit \cite{Corral03}.
Ref.\cite{SaichevSor06,SaichevSor07} found that expression (\ref{eq_10})
can fit rather well the empirical distributions of inter-event times, so as
to even improve on Corral's fit for short time scales, with $\alpha \approx 0.7-0.9$
and $n \approx 0.8-1.0$. These rather large intervals reflect
a corresponding insensitivity of the quantitative shape of $f(x)$ with respect
to $\alpha$ and $n$.
An analysis of the impact of the first nonlinear term
$\beta z^\gamma$ in the expansion (\ref{eq_3}) of $\Psi(z)$ in  the nonlinear equations
(\ref{eq_1}) and (\ref{eq_2}) led Saichev and Sornette to expect
``weak departures from the results obtained with the linear
approximation''. They added ``It thus appears that the statistics of
recurrence times is not sensitive enough to reveal the
importance of these nonlinear corrections which describe
the effect of cascades of generations of aftershocks.''
It turns out that this statement was premature, as shown by our full treatment
of the nonlinear equations. In particular, we identify significant biases
in the estimation of the parameter $n$ when using the linear approximation.
The reason lies in the fact that, for $\alpha$ close to $1$ (specifically
$1 \leq \gamma \leq 1.2$) as found in
Ref.\cite{SaichevSor06,SaichevSor07}, the first nonlinear correction
$\beta z^\gamma$ is only weakly  nonlinear. We show below that the inclusion
of the next term $\sim z^2$ changes somewhat the conclusions.
In contrast, the higher-order terms beyond $z^2$ do not change the conclusions.

\section{Analysis of the full nonlinear equations (\ref{eq_1}) and (\ref{eq_2})}

\subsection{Preparation of the equations and notations}

Using the first four summands of the expansion (\ref{eq_3}),
equations~(\ref{eq_1},\ref{eq_2}) can be written in the following form:
\begin{equation}
N_-(\tau,m_d)=\frac{Q(m_d)}{N_-(m_d)}+\delta \Phi(\tau)\otimes
N_-(\tau,m_d)+\sigma N_-(m_d)[\Phi(\tau)\otimes
N_-(\tau,m_d)]^2,\label{eq_11}
\end{equation}
\begin{equation}
\begin{array}{c}\displaystyle
N(t,\tau,m_d)=n[\Phi(t)\otimes N(t,\tau,m_d)+\Phi(t+\tau)\otimes
N_-(\tau,m_d)]-\\\displaystyle -\beta[\Phi(t)\otimes
N(t,\tau,m_d)+\Phi(t+\tau)\otimes
N_-(\tau,m_d)]^\gamma+\\\displaystyle +\eta[\Phi(t)\otimes
N(t,\tau,m_d)+\Phi(t+\tau)\otimes N_-(\tau,m_d)]^2,
\end{array}\label{eq_12}
\end{equation}
where
\begin{equation}
\sigma=\eta[1-Q^{1-2/\gamma}(m_d)].
\end{equation}
In the present case, $N_-(m_d)$ defined in (\ref{eq_5}) is not identical to $Q(m_d)/(1-\delta)$ as
in the linear approximation. Instead, $N_-(m_d)$ is the root of a simple quadratic equation.
However, the difference can be small: for instance, for $m_d-m_0=2,
\theta=0.03, \gamma=1.2, n=0.9$, we have
$Q(m_d)/(1-\delta) \simeq 1.93*10^{-2}$ compared with $N_-(m_d)\simeq 1.90*10^{-2}$.

Defining the dimensionless variables
$x=\lambda t$, $y=\lambda\tau$ with $\lambda=\omega
N_-(m_d)$ and the functions
\begin{equation}
M_-(y)=N_-\left(\frac y\lambda,m_d \right),\quad M(x,y)=N\left(\frac x\lambda,\frac
y\lambda,m_d\right),
\end{equation}
\begin{equation}
\Phi_\epsilon(x)=\lambda\Phi\left(\frac x\lambda\right).
\end{equation}
the equations (\ref{eq_11}) and (\ref{eq_12}) become
\begin{equation}
M_-(y)=\frac{Q(m_d)}{\lambda}+\delta \Phi_\epsilon(y)\otimes
M_-(y)+\sigma\lambda [\Phi_\epsilon(y)\otimes
M_-(y)]^2,\label{eq_13}
\end{equation}
\begin{equation}
\begin{array}{c}\displaystyle
M(x,y)=n[\Phi_\epsilon(x)\otimes M(x,y)+\Phi_\epsilon(x+y)\otimes
M_-(y)]-\\\displaystyle -\beta[\Phi_\epsilon(x)\otimes
M(x,y)+\Phi_\epsilon(x+y)\otimes M_-(y)]^\gamma+\\\displaystyle
+\eta[\Phi_\epsilon(x)\otimes M(x,y)+\Phi_\epsilon(x+y)\otimes
M_-(y)]^2~.
\end{array}\label{eq_14}
\end{equation}
For the numerical calculations, we transform equation (\ref{eq_13}) into
an equation for the new function $g(y)=1-M_-(y)/\lambda$:
\begin{equation}
\begin{array}{c}\displaystyle
g(y)=1-\delta-\frac{Q(m_d)}{\lambda}
+\delta\left[a_\epsilon(y)+\Phi_\epsilon(y)\otimes
g(y)\right]-\\\displaystyle
-\sigma\lambda\left[a_\epsilon(y)+\Phi_\epsilon(y)\otimes
g(y)-1\right]^2,\quad a_\epsilon(y)=\int\limits_x^\infty
\Phi_\epsilon(x')dx'~.
\end{array}\label{eq_15}
\end{equation}
Solving for $g(y)$ instead of $M_-(y)$ is more efficient numerically
because $g(y)$ is a monotonically decreasing function unlike
$M_-(y)$. This ensures a faster numerical convergence and a weaker
sensitivity to the finite mesh size of the discretization scheme.
Equations~(\ref{eq_14}) and (\ref{eq_15}) form the basis
for our numerical calculations.

\subsection{Numerical solution}

The first step is to solve (30) for the function $g(y)$ that we 
reformulate as equation  (\ref{eq_15a}) given in the Appendix A.
we use the method of successive approximations
to obtain the value of the function $g(y_i)$
on a regular grid $y_i=y_0+i\cdot dy$ with a small mesh $dy$.
The performance of this method is discussed in Appendix A
in the context of the linear approximation.

\begin{figure}[t]
\includegraphics[width=12cm]{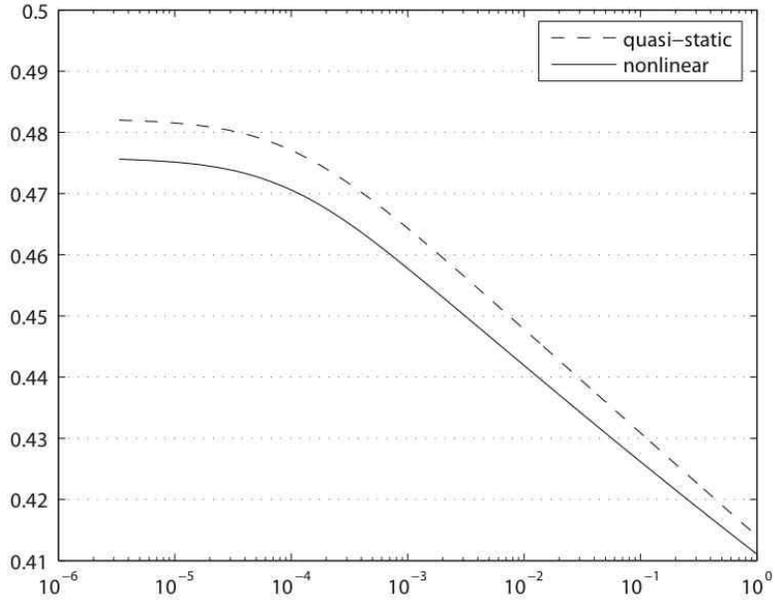}
\caption{Solution $g(y)$ of the nonlinear equation~(\ref{eq_15})
(continuous line) and quasi-static approximation~(\ref{eq_q}) (dashed line).
The parameters of the ETAS model are $m_d-m_0=2, \gamma=1.2, n=0.9, \theta=0.03$.}
\label{fig:quaziNL}
\end{figure}

\begin{figure}[t]
\includegraphics[width=12cm]{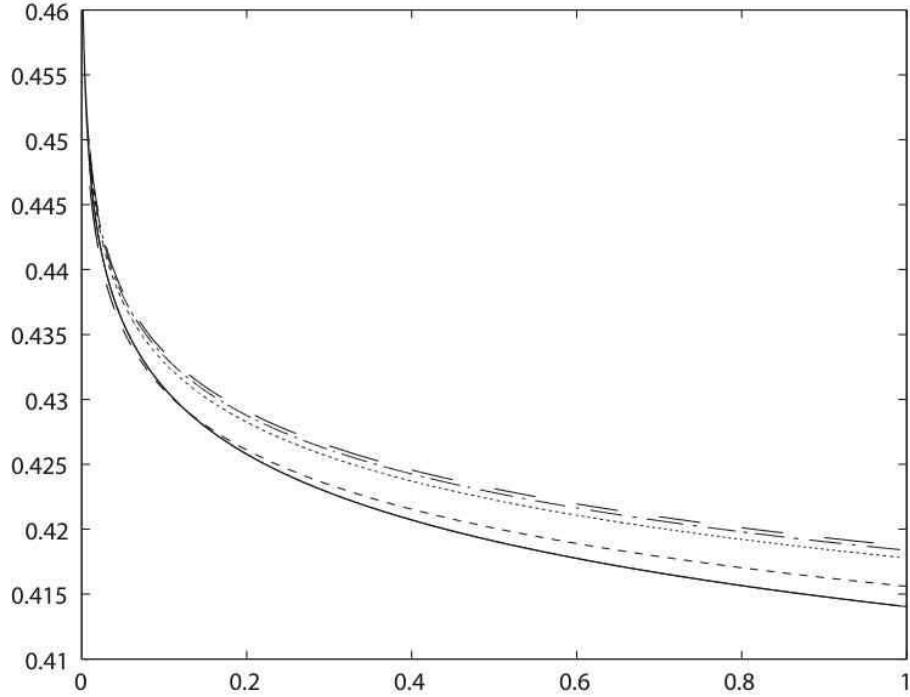}
\caption{Function $g(y)$ obtained with different schemes: (solid) -
quasi-static approximation developed in
\cite{SaichevSor06,SaichevSor07}; numerical solution of equation
(\ref{eq_15a}) for (dash-dotted) $dy= 10^{-5}$ and (long dashed) -
$10^{-4}$; numerical of solution of (\ref{eq_13}) for $M_-(y)$
yielding $g(y)=1-M_-(y)$: (dotted) $dy= 10^{-5}$ and (dashed)
$10^{-4}$. The ETAS parameters are $m_d-m_0=2, \gamma=1.2, n=0.9,
\theta=0.03$.} \label{fig:QuaNg}
\end{figure}

Fig.\ref{fig:quaziNL} shows the difference between the quasi-static
approximation~(\ref{eq_q}) and the solution of the nonlinear
equation~(\ref{eq_15}). One can observe that the nonlinear solution lies under the
quasi-static approximation, i.e., it gives a correction which is in
the opposite direction compared with the linear solution (see
fig.\ref{fig:QuaNg} ). 

The next step is to determine the function $M(x,y)$, obtained as the solution of
equation~(\ref{eq_14}). Note that the convolution operation involving $M_-(y)$
can now be expressed in terms of the known function $g(y)$:
\begin{equation}
\Phi_\epsilon(x+y)\otimes M_-(y)=
a_\epsilon(t)-a_\epsilon(t+\tau)- \Phi_\epsilon(x+y)\otimes g(y)~.
\label{eq_17}
\end{equation}
In the nonlinear case, the function $M(x,y)$ cannot be represented
analytically through the function $M_-(y)$ as in the linear case.
This means that we have to calculate $M(x,y)$, a function of
two variables (which significantly slows down the
calculation speed). Equation~(\ref{eq_17}) implies that the functions
$\Phi_\epsilon(x+y)$ and $g(y)$ should be estimated on the same grid points
$(x+y)_k$ and $y_l$. Therefore, the mesh sizes of $x$ and of
$y$ should be identical: $dx=dy$.

In order to determine the probability $P(\tau)=\varphi(y,m_d)$, we must also
estimate the integral
\begin{equation}
\int\limits_0^\infty M(x,y)dy~.
\end{equation}
This requires to span a large set of $y$ values in order to
approximate the theoretical one $[0; +\infty]$ which, together with the
condition $dx=dy$, make the problem very demanding in memory capacity.
For example, for  $x\in[0;1]$ and
$y\in[0;1]$ with $dy=dx\sim 10^{-4}$, $M(x,y)$ is a matrix with
$10^8$ elements. To alleviate this burden on memory capacity, we divide
the $y$-interval into smaller intervals $y_n, n=1,2,...,N$ and we
determine the matrix $M(x,y_n)$ consecutively for each of these sub-intervals.
Having determined the  probability function $\varphi(y,m_d)$ on each such small intervals
$y_n$, we use a simple smoothing polynomial interpolating scheme in order to prevent jumps
in its second order derivative.

\begin{figure}[t]
\includegraphics[width=12cm]{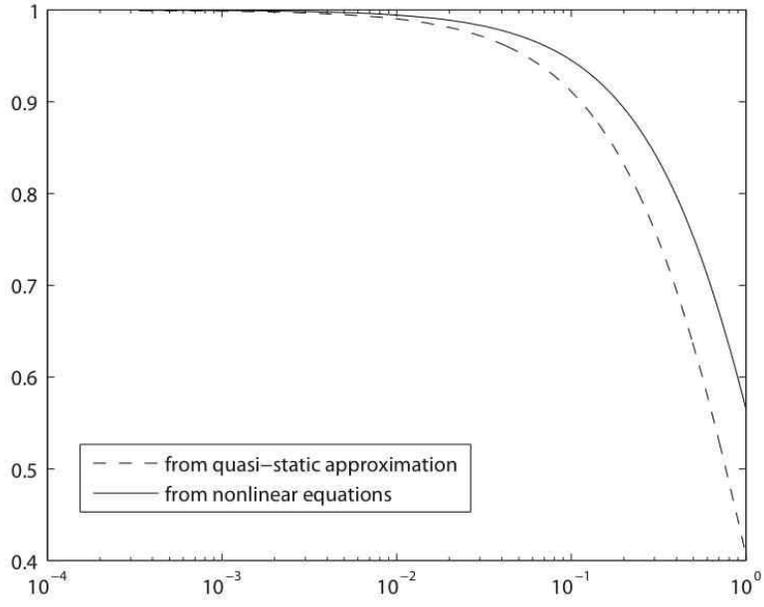}
\caption{Probability $P(y)$ defined in (\ref{eq_4}) obtained by the numerical
solution of the nonlinear equations (\ref{eq_14}) and (\ref{eq_15}) (continuous line)
compared with the quasi-static approximation reported
in Ref.~\cite{SaichevSor06,SaichevSor07} (dashed line). The ETAS parameters
are $m_d-m_0=2, \gamma=1.2, n=0.9, \theta=0.03$.
}\label{fig:prob}
\end{figure}

An example of the resulting probability $P(y)$ defined in (\ref{eq_4}) is shown
in fig.~\ref{fig:prob}, which identifies a significant difference between
the quasi-static approximation presented in Ref.~\cite{SaichevSor06,SaichevSor07}
and the nonlinear solution.

\subsection{Comparison between the linear and nonlinear versions of the theory and direct ETAS simulated catalogs}

We present a comparison for the pdf of inter-event times
obtained with
\begin{enumerate}
\item[(i)] the linear analytic quasi-static approximation,
\item[(ii)] the numerical solution of the nonlinear equations (\ref{eq_14}) and (\ref{eq_15})
and
\item[(iii)]  ``exact'' synthetic catalogs.
\end{enumerate}
The two former solutions are obtained by taking the
second order derivative of the functions $P(y)$ ($\varphi(y,m)$) shown in
 fig.~\ref{fig:prob}, according to (\ref{eq_deff}). The synthetic catalog was
 obtained using the method described in Appendix B. The ETAS parameters
 used here are: $n=0.9, \theta=0.05, \gamma=1.1, m_d-m_0=0$.

\begin{figure}[t]
\includegraphics[width=15cm]{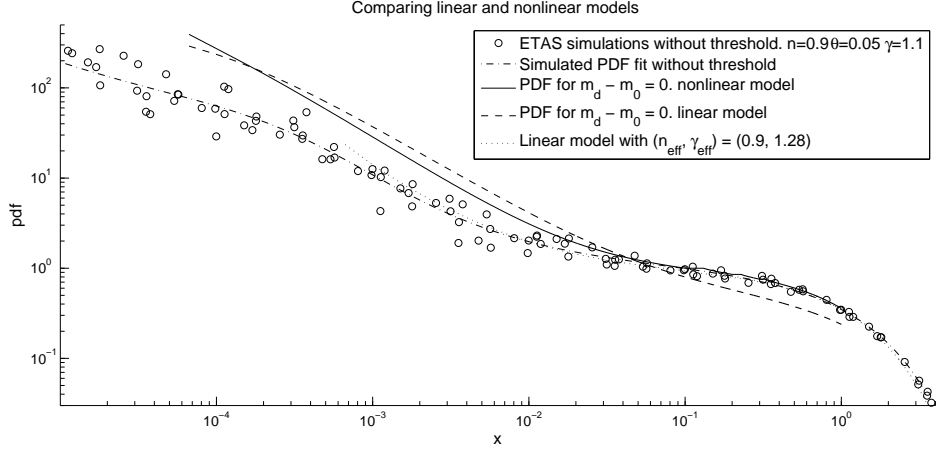}
\caption{Pdf of dimensionless inter-event times for the ETAS
parameters equal to $n=0.9, \theta=0.05, \gamma=1.1, m_d-m_0=0$.
The pdf obtained with the nonlinear theory leading to the
equations (\ref{eq_14}) and (\ref{eq_15}) is shown with the solid
line. The pdf obtained with the linear theory is shown as the
dashed line. The ``exact'' pdf obtained by the simulation method
described in Appendix B is shown in histogram form (circles) and
with a smoothing fit (dashed-dotted) performed with a function
defined as the ratio of a polynomial function of $5$-th order over
another polynomial function of $4$-th order, in terms of the
logarithm of the dimensionless inter-event times. The dotted line
shows the pdf obtained with the linear theory with effective
parameters $n_{\rm eff} = 0.86$ and $\gamma_{\rm eff} = 1.28$,
chosen to fit the ``exact'' pdf in the region $x \geq 4\cdot
10^{-2}$ where the nonlinear theory is performing well. }
\label{fig:lin_vs_nl}
\end{figure}

 Fig.~\ref{fig:lin_vs_nl} shows the three pdf's obtained by the three methods.
 For the ``exact'' pdf reconstructed from a synthetic ETAS catalog, we show
 both the histogram and a fit using a function constructed as the ratio between a
 polynomial function of order $5$ divided by another polynomial function
 of order $4$. These functions are expressed in terms of the logarithm of
 the dimensionless inter-event time. This fit has no pretence of rigor, it
 only provides a useful guide to the eye.

Fig.\ref{fig:lin_vs_nl} shows that the linear theory is
significantly in error while the nonlinear theory provides an
excellent agreement with the ``exact'' pdf  for values of the
dimensionless time interval $x \geq 4\cdot 10^{-2}$. For smaller
$x$'s, the difference is due to numerical errors in the treatment
of equations (\ref{eq_14}) and (\ref{eq_15}), which can be removed
by using an adaptive mesh size, as discussed shortly below. The
discrepancy between the ``exact'' pdf and the one obtained using
the  linear approximation implies that a fit of empirical pdf's
using the linear theory will likely provide spurious values for
the significant parameters $n$ and $\gamma$. Indeed, a good fit of
the ``exact'' synthetic pdf shown in fig.\ref{fig:lin_vs_nl} is
obtained with the linear theory using effective parameters $n_{\rm
eff} = 0.86$ and $\gamma_{\rm eff}=1.28$, showing here a
systematic bias of $5\%$ in the determination of $n$ and over
$16\%$ in the determination of $\gamma$ (and therefore of the
productivity $\alpha$).

\begin{figure}[t]
\includegraphics[width=15cm]{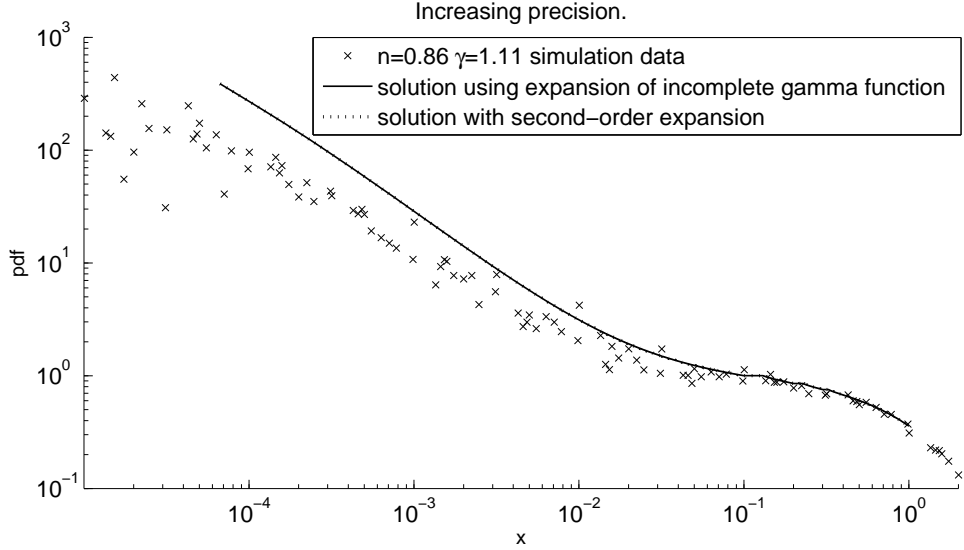}
\caption{Test showing that using the exact values of the
incomplete gamma function (\ref{eq_psi}) in the calculation of
equation (\ref{eq_14}) is undistinguishable from the solution
obtained with the expansion (\ref{eq_3}) which includes all four
terms up to second-order.} \label{fig:increase1}
\end{figure}

Let us now return to the discrepancy between the nonlinear theory
and the ``exact'' pdf observed for $x < 4\cdot 10^{-2}$ in fig.\ref{fig:lin_vs_nl}.
Two possible factors need to be discussed:
\begin{enumerate}
\item impact of terms of order higher than $z^2$ in the expansion
(\ref{eq_3}) of the function $\Psi(z)$ given by (\ref{eq_psi});
\item lack of convergence of the numerical scheme to solve equations
(\ref{eq_14}) and (\ref{eq_15}), due to a too large mesh size.
\end{enumerate}
Fig.~\ref{fig:increase1} rules out the first explanation, since
the solution of the nonlinear equations obtained by using the full expression
(\ref{eq_psi}) in the calculation of equation (\ref{eq_14}) is
undistinguishable from the solution obtained with the expansion
(\ref{eq_3}). This check and other tests confirm that there
is no need to complicate the computations by adding the calculation of
the incomplete gamma function. This is important when using
our theory for inverting the parameters from fits to empirical data, for instance.

\begin{figure}[t]
\includegraphics[width=15cm]{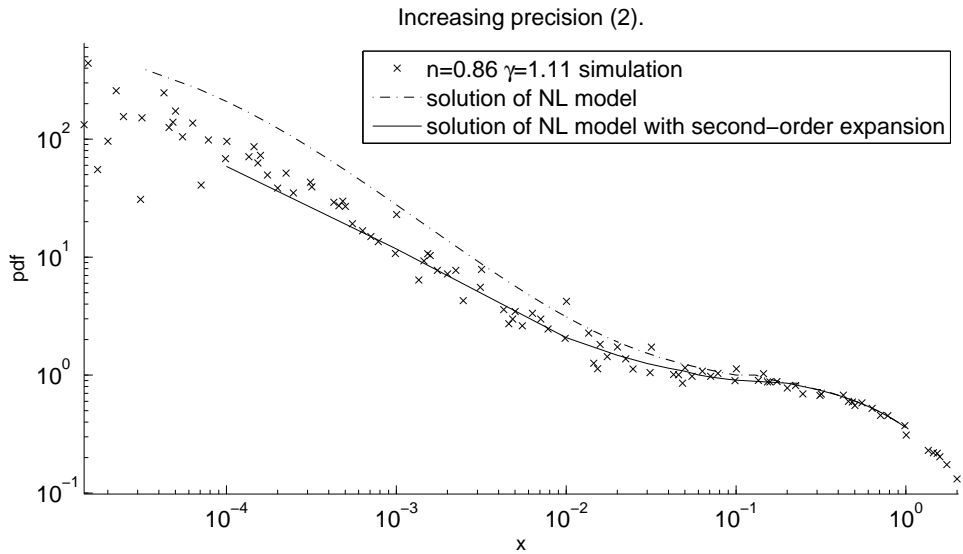}
\caption{Comparison between the  ``exact'' pdf obtained
from a synthetic ETAS catalog (crosses) with $n=0.86, \theta=0.05,
\gamma=1.11$ and the nonlinear theory without (dashed line) and with (continuous
line) adaptive mesh grid size as described in the text.}
\label{fig:increase2}
\end{figure}

With respect to the second factor, we improve the numerical precision
by varying the mesh size $dx$ so that $dx/x$ remains approximately
equal to $10^{-4}$ for $x<0.1$ while $dx$ is fixed at $10^{-4}$ for $x \geq 0.1$.
Thus, for $x \simeq 0.001$, we have chosen $dx \simeq 10^{-7}$, which is the limit
that we have been able handle due to limited numerical precision of the computer.
Fig~\ref{fig:increase2} shows for the example $n=0.86, \theta=0.05,
\gamma=1.11$ that the problem previously noted in fig.~\ref{fig:lin_vs_nl}
disappears: there is a good agreement between the ``exact'' pdf obtained
from the synthetic ETAS catalog and the nonlinear theory down to $x=0.001$.

\subsection{Test of the inversion of the parameters
$n$ and $\gamma$ using the nonlinear theory from a synthetic ETAS catalog}

\begin{figure}[t]
\includegraphics[width=15cm]{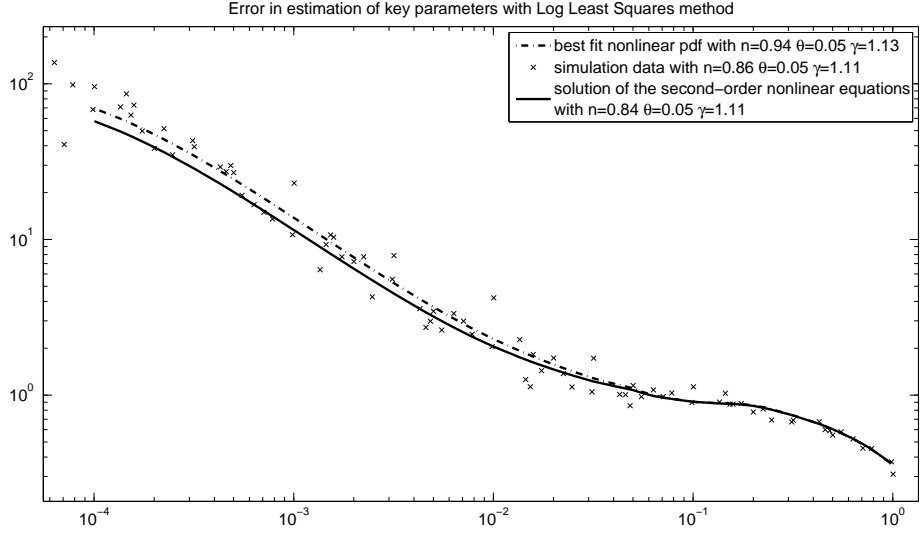}
\caption{Comparison between (i) the ``exact'' pdf of inter-event times (crosses)
 generated with the ETAS model according  to the method described in Appendix B with
$n_{\rm cat}=0.86, \theta_{\rm cat}=0.05, \gamma_{\rm cat}=1.11$
and $\{m_d-m_0\}_{\rm cat}=0$, which mimics a real-life situation
containing fluctuations, (ii) the best fit (dashed line)
with the nonlinear theory and (iii) the pdf obtained with the
nonlinear theory with the true values of the parameters
(continuous line). For simplicity, we impose the true values
$\theta=\theta_{\rm cat}=0.05$ and $m_d-m_0=\{m_d-m_0\}_{\rm
cat}=0$ in the best fit and invert for the two other parameters,
which yields $n_{\rm best~fit}=0.94 \pm 0.02$ and $\gamma_{\rm
best~fit}=1.13 \pm 0.02$.} \label{fig:sim_vs_nl}
\end{figure}

Consider a synthetic ETAS catalog of inter-event times
for some fixed values of the parameters $n_{\rm cat}, \theta_{\rm cat},
\gamma_{\rm cat}$  and $\{m_d-m_0\}_{\rm cat}$. In this example, we
take specifically $n_{\rm cat}=0.86, \theta_{\rm cat}=0.05, \gamma_{\rm cat}=1.11$
and $\{m_d-m_0\}_{\rm cat}=0$. Figure~\ref{fig:sim_vs_nl} shows the ``exact'' pdf of inter-event times (crosses), which mimics a real-life situation with statistical fluctuations.
In a real-life experiment, one would like to use the nonlinear theory to invert for the
unknown parameters $n, \theta, \gamma, m_d-m_0$.
In this goal, using the nonlinear theory, we calculate the predicted pdf
$f_{\rm NL}(x)$ of inter-event times
for fixed values of the parameters $n, \theta, \gamma$ and $m_d-m_0$.
For a given set of these four parameters, we construct the mean-square error
of the logarithm of the pdf over the $N$ inter-event times of the catalog:
\begin{equation}
LLS(n, \theta, \gamma, m_d-m_0)=\sum_{i=1}^N \left(\ln f_{\rm NL}(x_i)-\ln f_{\rm cat}(x_i)\right)^2~,
\label{eq:lls}
\end{equation}
where $f_{\rm NL}(x_i)$ is the predicted pdf at the dimensionless inter-event time $x_i$
given by the nonlinear theory and $f_{\rm cat}(x_i)$ is the corresponding empirical
pdf (in the synthetic catalog). $LLS(n, \theta, \gamma, m_d-m_0)$ quantifies how well
the nonlinear prediction for the pdf of inter-event times can describe the (synthetic) data.
The unknown parameters $n, \theta, \gamma, m_d-m_0$ are then obtained by finding the quadruplets which makes $LLS(n, \theta, \gamma, m_d-m_0)$ minimum.

In practice, given the computational cost of the numerical
solution of the nonlinear theory, we have found unpractical to
explore systematically the four dimensional parameter space (with
super-computer resources, this is not excluded but the next
section removes the motivation to explore further this option as
we will see). For the sake of demonstration, we assume that we
already know $\theta=\theta_{\rm cat}=0.05$ and
$m_d-m_0=\{m_d-m_0\}_{\rm cat}=0$. We are then left with searching
for the remaining parameters $n$ and $\gamma$. For this, we form a
grid in the $(n, \gamma)$ plane over which we find the minimum of
$LLS(n, \gamma; \theta_{\rm cat}, \theta_{\rm cat})$. The
corresponding inverted values are $n_{\rm best~fit}=0.94 \pm 0.02$
and $\gamma_{\rm best~fit}=1.13 \pm 0.02$. The recovery of $\gamma$
(and therefore of the productivity exponent $\alpha$) is good, while
there is $9\%$ error on $n$. Figure~\ref{fig:sim_vs_nl} shows that this best pdf fits
well the ``exact'' pdf obtained from the ETAS catalog and is not
far from the pdf predicted by the nonlinear theory with the true
parameters.

\section{On the lack of power of the pdf of inter-event times to invert for
the clustering parameter $n$ and other parameters}

\begin{figure}[t]
\includegraphics[width=15cm]{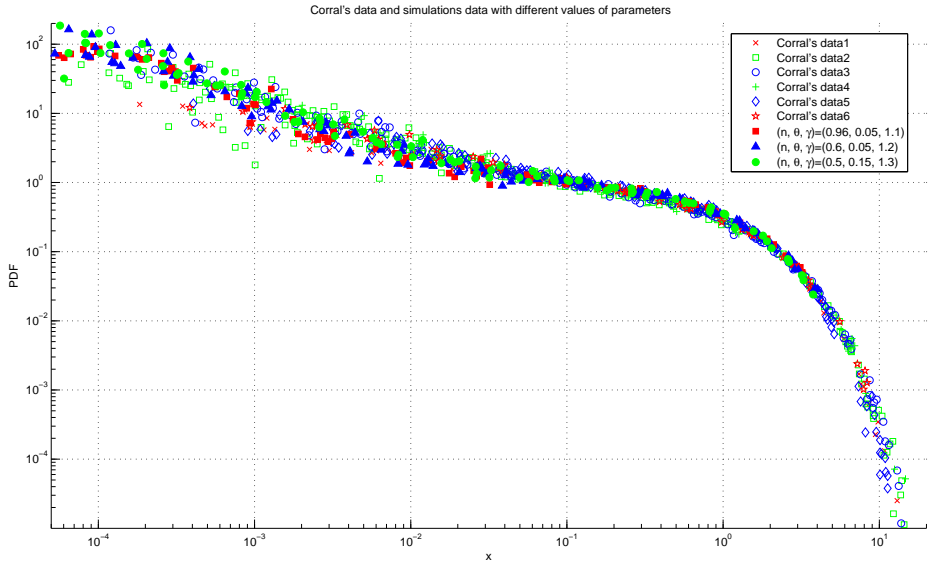}
\caption{Comparison between the pdf's of inter-event times in synthetic ETAS catalogs with three
different sets of parameters and the pdf's obtained by Corral
in different regions of the world.}
\label{fig:simcor}
\end{figure}

The title of this section is motivated by fig.\ref{fig:simcor} comparing
the pdf of inter-event times in synthetic ETAS catalogs with three
different sets of parameters and the pdf's obtained by Corral
in different regions of the world \cite{Corral03}.

First, one can observe that the three triplets $(n=0.96, \theta=0.05, \gamma=1.1)$,
$(n=0.6, \theta=0.05, \gamma=1.2)$ and $(n=0.5, \theta=0.15, \gamma=1.1)$
give almost the same pdf's over the whole range of dimensionless inter-event times
$10^{-4} \leq x \leq 15$. For $x>0.1$, the data collapse is almost perfect, while the
scatter is larger for the smaller $x$ values.
This is bad news for the determination of the
clustering parameter $n$ in particular, since relatively small changes in the
Omori law parameter $\theta$ and in the productivity law parameter $\gamma$ can compensate for a quite significant change in the branching ratio $n$. This suggests that previous
claims on the use of the pdf of inter-event times to extract efficiently the
clustering parameter have been over-optimistic \cite{Hainzl2006,Hardebeck07}.

Second, fig.\ref{fig:simcor} shows that the three chosen triplets of parameters
are basically equally good at fitting Corral's data sets \cite{Corral03}. We note
 that, again for $x>0.1$, all empirical data and ETAS simulations present
an almost perfect collapse on a quasi-universal curve. Larger scatter
characterize smaller $x$ values, which is the region to scrutinize in the
hope of extracting some useful constraints on the parameter values.

\begin{figure}[t]
\includegraphics[width=15cm]{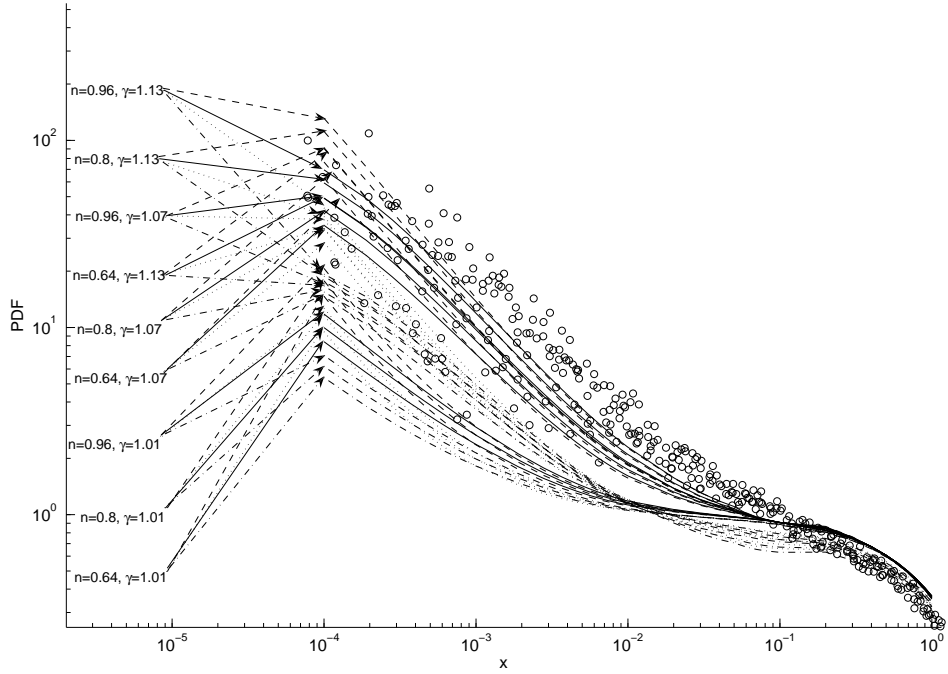}
\caption{Comparison between the pdf's obtained by Corral
in different regions of the world and the 36 pdf's of inter-event times calculated
with the nonlinear theory for all combinations of the following sets (with
a fixed $\theta = 0.05$): 
$n=0.64, 0.8, 0.96$; (ii) $\gamma=1.01,
1.07, 1.13$; (iii) $m_d-m_0=0.1, 1$; (iv) $\epsilon(c)=10^{-4}, 10^{-5}$. 
Solid lines:   $m_d-m_0=0.1,\;c=10^{-4}$;
dashed lines: $m_d-m_0=0.1,\;c=10^{-5}$;
dashed-dotted lines: $m_d-m_0=1,\;c=10^{-4}$;
dotted lines: $m_d-m_0=1,\;c=10^{-5}$. }
 \label{fig:cm_test}
\end{figure}

Actually, the situation is even more involved since, in addition to the
parameters $n, \theta$ and $\gamma$, a genuine inversion needs
also to determine $m_d-m_0$ (whose impact is significant as shown
in Ref.~\cite{SorWerner05}) as well as the regularizing constant $c$
in the Omori law (\ref{eq_o}). Fig.~\ref{fig:cm_test} presents the pdf's calculated
with the nonlinear theory for different sets of four of these parameters
$(n, \gamma, m_d-m_0, c)$ with a fixed $\theta=0.05$, together with Corral's data. 
For the pdf's obtained from the nonlinear theory, we used all combinations
between the three values $n=0.64, 0.8, 0.96$, the three values
$\gamma=1.01, 1.07, 1.13$, two values $m_d-m_0=0.1, 1$ and two values
$\epsilon(c) \equiv \lambda c=10^{-4}, 10^{-5}$, corresponding to a total of
36 combinations. One can observe roughly two clusters among these 36 
theoretical curves. All curves with $\gamma=1.01$ belong to the lower cluster,
which is clearly not fitting the data. The upper cluster, which is in better
agreement with the data, corresponds to the larger values $\gamma=1.07$ and 
$1.13$. This suggests that the productivity parameter $\alpha$ is
likely to be smaller than (instead of equal to) the $b$-value of the Gutenberg-Richter law 
(recall that $\gamma = b/\alpha$). There is also a smaller impact of 
 $\epsilon$ and of $m_d-m_0$: in general, higher values of these parameters
displace the pdf downward.

The comparison between these 36 theoretical pdf's and Corral's data
in figure \ref{fig:cm_test} shows that there are large uncertainties
in the inversion of the parameters. One could argue that the parameter
$\theta$ should perhaps be modified to a value different from $0.05$ in order
to better describe the data for small $x$'s. But this region is very sensitive
to errors such as resulting from incompleteness \cite{HelmKJ,Kagan03}
and its use is problematic.

\section{Conclusion}

We can conclude by
the following rather conservative assessment. 
Recalling the definition of the dimensionless variable
$x=\lambda\tau$,  where $\lambda$ is the average seismicity rate
of a given region and $\tau$ is a realization of the random variable
defined as the inter-event time between two successive events in that region,
we observe on the one-hand that the
range of dimensionless inter-event times $x \geq 0.1$ is probably quite
reliable from an empirical view point but the corresponding pdf's are
remarkably insensitive to the specific values of the clustering parameter,
Omori law exponent, productivity exponent and completeness of the catalogs.
On the other hand, the range of $x<0.1$ which would promise to give
more sensitivity is not only highly unreliable but also lacks significant
power to obtain a good inversion due to the existence of many
almost equally good fits with quite different sets of parameters.
Our theoretical analysis and its comparison with Corral's data does not
seem to support the proposition that inter-event time distributions could
provide a new and more reliable way to measure of the so-called background earthquake activity
as suggested in Ref.\cite{Hainzl2006,Hardebeck07}. 

{\bf Acknowledgements}: We are grateful to A. Corral for sharing his data with us.

\pagebreak

\section*{Appendix A: Numerical solution for the linear approximation}

This appendix provides a validation step of the numerical discretization scheme
that we have developed to solve the nonlinear equations~(\ref{eq_14}) and (\ref{eq_15}).
Here, we apply this scheme to the linear approximation and compare the result with those
which are available analytically. In the linear case, the equation (\ref{eq_15}) for $g(y)$
reduces to the following implicit linear integral equation
\begin{equation}
g(y)=\delta [a_\epsilon(y)+\Phi_\epsilon (y)\otimes g(y)]~,
\label{eq_15a}
\end{equation}
where $\delta$ is defined in (\ref{eq_6}).
To solve (\ref{eq_15a}), we use the method of successive approximations
to obtain the value of the function $g(y_i)$
on a regular grid $y_i=y_0+i\cdot dy$ with a small mesh $dy$. This method
is adapted to the treatment of the convolution integral in the
right-hand-side of~(\ref{eq_15a}). This simple method is fast and provides
good convergence. For example, with $dy\sim 10^{-4}$, the calculation
converges on the 15-th iteration with a residual absolute error
$\sim 10^{-15}$.

\begin{figure}[t]
\includegraphics[width=12cm]{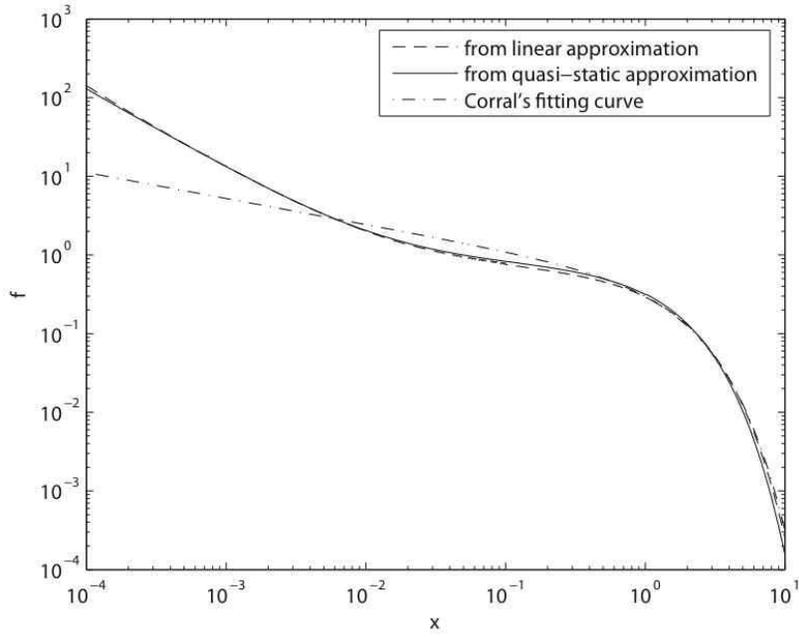}
\caption{Probability density functions (pdf) of inter-event times
obtained by using the function  $g(y)$ solution of (\ref{eq_15a})
with $dy= 10^{-5}$, with (\ref{eq_16}) and  (\ref{eq_deff}) (continuous line).
The pdf obtained by Saichev and Sornette with the quasi-static approximation \cite{SaichevSor06,SaichevSor07} is shown as the dashed line. Corral's fitting curve
is the dotted-dashed line. The parameters of the ETAS model are
$m_d-m_0=2, \gamma=1.2, n=0.9, \theta=0.03$.}
\label{fig:pdflin}
\end{figure}

This is illustrated in Fig.~\ref{fig:pdflin} which shows the function $g(y)$
in the linear approximation, obtained directly from the numerical
solution of equation (\ref{eq_15a}) for  $dy= 10^{-5}$ and $10^{-4}$, and by using
the equation for $M_-(y)$ for  $dy= 10^{-5}$ and $10^{-4}$. As
mentioned in the main text, the convergence is faster when using $g(y)$
compared with using $M_-(y)$. Eventually, as the mesh size goes to zero,
both methods converge towards the same estimation. An illustration of this
convergence is given with $dy= 10^{-5}$, which shows much better agreement
between the two estimations compared with the results obtained for $dy= 10^{-4}$.
The function $g(y)$ obtained by solving (\ref{eq_15a}) is
almost identical for  $dy= 10^{-5}$ and $10^{-4}$, demonstrating the faster convergence
of this scheme. Fig.~\ref{fig:pdflin} also shows that the
quasi-static approximation is not perfect, but exhibits a relative error
of about no more than $1\%$ in this example. Once $g(y)$ has been obtained,
the distribution of inter-event times is obtained from equation
(\ref{eq_9}), which can be expressed here as
\begin{equation}
\varphi(y,m_d)=\exp\left(-\frac{1-n}{1-\delta}y-
\frac{1-n}{\delta}\Delta\int\limits_0^y g(x)dx \right)~,
\label{eq_16}
\end{equation}
and with (\ref{eq_deff}). Using the function  $g(y)$ obtained by solving (\ref{eq_15a})
with $dy= 10^{-5}$ gives the dimensionless pdf of inter-event times shown in
Fig.~\ref{fig:pdflin}. For comparison is also shown the pdf obtained by Saichev
and Sornette with the quasi-static approximation \cite{SaichevSor06,SaichevSor07}. There is an excellent agreement between the two methods.

\pagebreak

\section*{Appendix B: ETAS simulations}

This appendix describes how we construct the pdf of inter-event times
in specific synthetic catalogs generated with the ETAS model. Actually, we
do not generate synthetic catalogs. Instead, we use the analytical form
of the cumulative distribution function (cdf) $F_{k+1}(\tau)$ of the waiting time $\tau$
between the $k$-th and  the $(k+1)$-th event, knowing the times
and magnitudes of the preceding $k$ events, to draw the occurrence time
of this $(k+1)$-th event. Generating in this way 1000 or more inter-event times,
we use logarithmic bins to construct the
histogram of these inter-event times. This construction provides the ``true'' or ``exact''
numerical benchmark against which to compare our theory and the empirical data.

The cdf $F_{k+1}(\tau)$ is obtained by recurrence as follows.
For the first event, $F_{1}(\tau)$ is nothing but the cumulative probability of
occurrence of a spontaneous (background) shock since
the origin of time, given by definition
by the Poisson law with rate $\omega$:
\begin{equation}
F_1(\tau)=1-e^{-\omega\tau}.\label{eq_spont}
\end{equation}

The cdf $F_2(\tau)$ of the waiting time from the first to the second shock is made
of two contributions: (i) the second shock may again be a background event
or (ii) it may be triggered by the first shock. This yields
\begin{equation}
F_2(\tau)=1-e^{-\omega\tau} e^{-\rho_1(1-a(\tau))}~,
\label{eq_second}
\end{equation}
where $\rho_1=\rho(m_1)$ is the productivity of the first shock obtained from
expression (\ref{producad}) given its magnitude $m_1$, and
$a(\tau)$ is defined in (\ref{at}).

All following shocks are similarly either a background event or triggered
by one of the preceding events. The cdf $F_3(\tau)$ of the waiting time between
the second and the third shocks is thus given by
\begin{equation}
F_3(\tau)=1-e^{-\omega\tau} e^{-\rho_1(a(\tau_2)-a(\tau_2+\tau))
-\rho_2(1-a(\tau))}~,
\end{equation}
where $\tau_2$ is the realized time interval between the first and the second
shocks. Iterating, we obtain the cdf $F_k(\tau)$ for the waiting between the
$(k-1)$-th and $k$-th shocks under the following form
\begin{equation}
F_k(\tau)=1-e^{-\omega\tau} \exp\left[-\sum_{i=1}^{k-1}
\rho_{k-i}\cdot \left(a\left(\sum_{j=2}^{i}\tau_j\right)
-a\left(\sum_{j=2}^{i}\tau_j+\tau\right)\right)\right]~,
\label{eq_k}
\end{equation}
where $\tau_j$ is the waiting time between the $(j-1)$-th event
and the $j$-th event, and $\rho_i=\rho(m_i)$ is the productivity of the $i$-th shock obtained from
expression (\ref{producad}) given its magnitude $m_i$.

In order to generate the $(k+1)$-th inter-event time interval
between the occurrence of the $k$-th and $(k+1)$-th shock, it is
necessary to know the $k$ previous inter-events times between the
$k$ previous shocks and their $k$ magnitudes. Since, in the ETAS
model, the magnitudes are drawn independently  according to the
Gutenberg-Richter distribution (\ref{eq_g}), they can be generated
once for all. In order to generate a catalog of $N$ events, we
thus draw $N$ magnitudes from the law (\ref{eq_g}). In order to
generate the corresponding $N$ inter-event times, we use the
expression (\ref{eq_k}) iteratively from $k=1$ to $k=N$ in a
standard way: since any cdf $F(x)$ of a random variable $x$  is by construction
itself uniformly distributed in $[0,1]$, we obtain a given realization $x*$
of the random variable $x$ by drawing a random number $r$ uniformly in $[0,1]$
and by solving the equation $F(x*)=r$. In our case,
we generate $N$ independent uniformly distributed random
numbers $x_1, ..., x_N$ in $[0,1]$ and determine each $\tau_i$
successively as the solution of $F_i(\tau_i)=x_i$.

\clearpage

%FIGURE 1

%FIGURE 2

%FIGURE 3

%FIGURE 4

%FIGURE 5

%FIGURE 6

%FIGURE 7

%FIGURE 8

%FIGURE 9

%FIGURE 10

%FIGURE11


\clearpage
\begin{thebibliography}{widest-label}

\bibitem{Baketal02} Bak, P., K. Christensen, L. Danon, and T. Scanlon (2002),
%Unified scaling law for earthquakes,
Phys. Rev. Lett., 88(17), 178501.

\bibitem{Barabasi_Nature05} Barab\'asi, A.-L., Nature 435, 207 (2005).

\bibitem{Corral03} Corral, A.,
%Local distributions and rate fluctuations in a unified scaling law for earthquakes,
Phys. Rev. E 68, 035102(R) (2003).

\bibitem{Corral2004a} Corral, A.,
%Long-term clustering, scaling, and universality in the temporal occurrence of earthquakes,
Phys. Rev. Lett. 92, 108501 (2004).

\bibitem{Corral2004b} Corral, A.,
%Universal local versus unified global scaling laws in the statistics of seismicity,
Physica A, 340, 590-597 (2004).

\bibitem{Corral2005a} Corral, A.,
%Mixing of rescaled data and bayesian inference for earthquake recurrence times,
Nonlinear Processes in Geophysics, 12, 89-100 (2005).

\bibitem{Corral2005b} Corral, A.,
%Renormalization-group transformations and correlations of seismicity,
Phys. Rev. Lett., 95, 028501 (2005).

\bibitem{Corral2006} Corral, A.,
%Universal earthquake-occurrence jumps, correlations with time, and anomalous diffusion,
Phys. Rev. Lett., 97, 178501 (2006).

\bibitem{Corral_Christensen06} Corral, A., and K. Christensen,
%Comment on Уearthquakes descaled: On waiting time distributions and scaling lawsУ,
Phys. Rev. Lett., 96, 109801 (2006).

\bibitem{CraneSornette07} Crane, R. and D. Sornette,
Searching with viral dynamics on social networks: Application to YouTube,
working paper (2007).

\bibitem{DavidsenGoltz04} Davidsen, J., and C. Goltz,
%Are seismic waiting time distributions universal?,
Geophys. Res. Lett., 31, doi:10.1029/2004GL020892 (2004).

\bibitem{Davisenetal07} Davidsen, J., S. Stanchits and G. Dresen,
%Scaling and universality in rock fracture
Phys. Rev. Lett. 98, 125502 (2007).

\bibitem{Gao} Gao, J., Y. Cao, and J. Hu,
Recurrence Time Distribution, Renyi Entropy, and Pattern Discovery,  2005 Conference on Information Sciences and Systems, The Johns Hopkins University, March 16Р18, 2005.

\bibitem{Hainzl2006} Hainzl, S.,  F. Scherbaum,  C. Beauval.
%Estimating Background Activity Based on Interevent-Time Distribution.
Bulletin of the Seismological Society of America 96 (1), 313-320, (2006).

\bibitem{Hardebeck07} Hardebeck, J.,
Background seismicity rates from interevent-time statistics: spatial patterns appear stationary through time, working paper (2007).

\bibitem{Hawkes71a} Hawkes, A. G.,
%Spectra of some self-exciting and mutually exciting point processes,
Biometrika, 58(1), 83-90 (1971a).

\bibitem{Hawkes71b} Hawkes, A. G.,
%Point spectra of some mutually exciting point processes,
J. Royal Stat. Soc. Series B (Meth.), 33(3), 438-443 (1971).

\bibitem{HawkesOakes74} Hawkes, A. G., and D. Oakes,
%A cluster process representation of a self-exciting process,
J. of Appl. Prob., 11(3), 493-503 (1974).

\bibitem{HelmKJ} Helmstetter, A., Y. Y. Kagan, and D. D. Jackson,
%Importance of small earthquakes for stress transfers and earthquake triggering,
J. Geophys. Res., 110, B05S08, doi:10.1029/2004JB003286 (2005).

\bibitem{Helmsor02} Helmstetter, A. and D. Sornette,
%Sub-critical and supercritical regimes in epidemic models of earthquake aftershocks,
J. Geophys. Res. 107, NO. B10, 2237, doi:10.1029/2001JB001580 (2002).

\bibitem{Helmsor03} Helmstetter, A. and D. Sornette,
%Importance of direct and indirect triggered seismicity in the ETAS model of seismicity,
Geophys. Res. Lett. 30 (11) doi:10.1029/2003GL017670 (2003).

\bibitem{Kagan99} Kagan, Y. Y.,
%Universality of the seismic moment-frequency relation,
Pure and Appl. Geophys., 155, 537-573 (1999).

\bibitem{Kagan03} Kagan, Y.Y.,
% Accuracy of Modern Global Catalogs,
Physics of the Earth and Planetary Inter. 135, 173-209 (2003).

\bibitem{KK81} Kagan, Y. Y., and L. Knopoff,
%Stochastic synthesis of earthquake catalogs,
J. Geophys. Res., 86 (B4), 2853-2862 (1981).

\bibitem{Lindmanetal05} Lindman, M., K. Jonsdottir, R. Roberts, B. Lund, and R. Bdvarsson,
%Earthquakes descaled: On waiting time distributions and scaling laws,
Phys. Rev. Lett., 94, 108501 (2005).

\bibitem{Lindmanetal06} Lindman, M., K. Jonsdottir, R. Roberts, B. Lund, and R. Bdvarsson,
%Reply to comment by A. Corral and K. Christensen,
Phys. Rev. Lett., 96, 109802 (2006).

\bibitem{Livina06} Livina, V. N., S. Havlin, and A. Bunde,
%Memory in the occurrence of earthquakes,
Phys. Rev. Lett., 95, 208501 (2006).

\bibitem{Livina} Livina, V., S. Tuzov, S. Havlin and A. Bunde,
%Recurrence intervals between earthquakes strongly depend on history,
Physica A 348, 591-595 (2005).

\bibitem{Molchan05} Molchan, G.,
%Interevent time distribution in seismicity: A theoretical approach,
Pure and Appl. Geophys., 162, 1135-1150 (2005).

\bibitem{Ogata88} Ogata, Y.,
%Statistical models for earthquake occurrence and residual analysis for point processes,
J. Am. Stat. Assoc., 83, 9-27 (1988).

\bibitem{Pisaetal07} Pisarenko,  V.F., A. Sornette, D. Sornette and M.V. Rodkin,
New Approach to the Characterization of Mmax and of the Tail of the Distribution
of Earthquake Magnitudes, in press in Pure and Applied Geophysics (2008)
(\url{http://arxiv.org/abs/physics/0703010})

\bibitem{Reasenberg85} Reasenberg, P.,
%Second-order moment of central California seismicity, 1969Р1982,
J. Geophys. Res. 90, 5479-5495 (2005).

\bibitem{SaichevSor06} Saichev, A. and D. Sornette,
%``Universal'' Distribution of Inter-Earthquake Times Explained,
Phys. Rev. Letts. 97, 078501 (2006).

\bibitem{SaichevSor07} Saichev, A. and D. Sornette,
%Theory of Earthquake Recurrence Times,
J. Geophys. Res., 112, B04313, doi:10.1029/2006JB004536 (2007).

\bibitem{Sornetteetal04} Sornette, D., F. Deschatres, T. Gilbert and Y. Ageon,
%Endogenous Versus Exogenous Shocks in Complex Networks: an Empirical Test Using Book Sale Ranking,
Phys. Rev. Letts. 93 (22), 228701 (2004).

\bibitem{SorWerner05} Sornette. D. and M.J. Werner,
%Apparent Clustering and Apparent Background Earthquakes Biased by Undetected Seismicity,
J. Geophys. Res.,КVol.К110 (B9),КB09303, 10.1029/2005JB003621 (2005).

\bibitem{Utsuetal95} Utsu, T., Y. Ogata, and R. S. Matsu'ura,
%The centenary of the Omori formula for a decay law of aftershock activity,
J. Phys. Earth, 43, 1-33 (1995).

\bibitem{Vasquez_et_al_06} Vazquez, A., J. G. Oliveira, Z. Dezso, K. I. Goh, I. Kondor, and A. L. Barabasi,
%Modeling bursts and heavy tails in human dynamics,
Physical Review E 73, 036127 (2006).

\bibitem{Zaslavsky91} Zaslavsky, G.M. and M.K. Tippett,
%Connection between recurrence-time statistics and anomalous transport,
Phys. Rev. Lett. 67, 3251-3254 (1991).

\end{thebibliography}
\end{document}